\documentclass[%
 reprint,
 amsmath,amssymb,
 aps,
]{revtex4-2}

\usepackage{graphicx}
\usepackage{dcolumn}
\usepackage{bm}
\usepackage{float}
\usepackage{hyperref}
\usepackage{subcaption}

\begin{document}


\title{Quantum entanglement provides a competitive advantage in adversarial games}

\author{Peiyong Wang}
 \email{Peiyong.Wang@csiro.au}
\author{Kieran Hymas}%
 \email{Kieran.Hymas@csiro.au}

\author{James Quach}
 \email{James.Quach@csiro.au}

\affiliation{%
 Technology, CSIRO Clayton, Research Way, Clayton, VIC 3168, Australia.
}%

\date{\today}

\begin{abstract}
Whether uniquely quantum resources confer advantages in fully classical, competitive environments remains an open question. Competitive zero-sum reinforcement learning is particularly challenging, as success requires modelling dynamic interactions between opposing agents rather than static state–action mappings. Here, we conduct a controlled study isolating the role of quantum entanglement in a quantum–classical hybrid agent trained on Pong, a competitive Markov game. An 8-qubit parameterised quantum circuit serves as a feature extractor within a proximal policy optimisation framework, allowing direct comparison between separable circuits and architectures incorporating fixed (CZ) or trainable (IsingZZ) entangling gates. Entangled circuits consistently outperform separable counterparts with comparable parameter counts and, in low-capacity regimes, match or exceed classical multilayer perceptron baselines. Representation similarity analysis further shows that entangled circuits learn structurally distinct features, consistent with improved modelling of interacting state variables. These findings establish entanglement as a functional resource for representation learning in competitive reinforcement learning.

\end{abstract}

\maketitle


\section{Introduction}

Competitive environments arise in many fields, from quantitative trading \cite{Treynor1981-co, Treynor1999-au} to modern reinforcement learning (RL) \cite{Silver2016-jm, Silver2017-db, Silver2017-hq}, where one player's gain is usually another's loss. In game theory, the simplest abstraction of this antagonism is the two-player zero-sum game, whose value under optimal play is characterised by the minimax theorem \cite{v-Neumann1928-ai}. Many competitive decision problems are sequential rather than one-shot, and can be formalised as (zero-sum) stochastic games—also known as Markov games—in which two agents repeatedly act in a shared state with stochastic transitions \cite{Shapley1953-ls, Littman1994-mx}. This perspective aligns naturally with modern RL, where self-play in competitive games has powered several landmark research \cite{Silver2016-jm, Silver2017-db, Silver2017-hq}. Pong provides a simple, controlled environment to explore modified learning dynamics in the presence of adversarial play. It is a classic two-player competitive game, and a common RL benchmark formulation assigns rewards that are largely anti-symmetric between the two sides (e.g., $\pm 1$ on scoring), making it a close approximation to a zero-sum Markov game \cite{ALEPaper, Mnih2013-by, Mnih2015-ip}. Recently, Pong has also been used as a platform to demonstrate learning in neuroplastic biological samples \cite{Kagan2022-rq}. These competitive benchmarks, therefore, offer a natural testbed for probing whether uniquely quantum resources—most notably entanglement—can yield measurable benefits for learning and decision making.

Quantum computing has long been anticipated to deliver improvements over classical algorithms across a range of domains, including optimisation, quantum system simulation, and machine learning, by offering advantages in computational complexity and predictive performance. A canonical example is Grover’s search algorithm, where quantum superposition enables an oracle to evaluate all database entries in parallel, yielding a quadratic speedup with $\mathcal{O}(\sqrt{N})$ time complexity \cite{grover1996fastquantummechanicalalgorithm}. Beyond superposition, other non-classical quantum resources have also been shown to provide advantages: quantum contextuality underpins the memory advantage of contextual recurrent neural networks (CRNNs) over classical recurrent neural networks (RNNs) \cite{Anschuetz2023-mk}. Among these resources, quantum entanglement is widely regarded as a key enabler of quantum advantage. Entanglement enables models and algorithms to exploit quantum correlations that lack classical counterparts, potentially enhancing the expressive power and learning efficiency of quantum machine learning (QML) architectures. To date, however, systematic empirical evidence for such benefits remains limited. Existing benchmarking studies have primarily focused on supervised learning, particularly (binary) classification tasks \cite{Bowles2024-kb}. Within this setting, the data reuploading model \cite{Perez-Salinas2020-gi}, also known as the quantum Fourier model (QFM) \cite{Schuld2021-kv}, stands out as one of the few quantum models for which entanglement has been shown to improve classification accuracy \cite{Bowles2024-kb}. These results suggest that entanglement can meaningfully enhance the performance of variational quantum models, motivating further investigation of its impact beyond supervised learning scenarios.

Research on quantum reinforcement learning (QRL) has established several forms of provable quantum advantage, but these proofs typically rely on settings where a quantum agent interacts with a quantum environment or communicates through a quantum channel. For example, \cite{Saggio2021-lb} demonstrates that access to quantum channels can yield algorithmic speed-ups for QRL agents. In Ref. \cite{Wu2025-hm}, the authors introduced a quantum deep deterministic policy gradient algorithm for the quantum state generation and the eigenvalue problems. Quantum signals can also be harnessed to enhance the average reward outcomes in the infinite-horizon Markov decision problem \cite{Ganguly2025-rr}. Such results, however, do not readily extend to classical RL environments, which are far more complex and less structured than the idealised models commonly assumed in theoretical quantum machine learning analyses. Most proofs of quantum advantage in machine learning rely on specific structures in the data that can be efficiently utilised by a quantum machine learning model, but not by a classical model. Popular choices for such structures are those related to cryptography with a (classical) hardness proof, such as the data constructed based on the decisional Diffie-Hellman assumption \cite{Sweke2021-uh}, those based on weak pseudorandom functions in \cite{Pirnay2023-oy}, and the classical data constructed based on the discrete logarithm algorithm in \cite{Liu2021-yj}. The hidden subgroup problem is also a popular choice for proving quantum advantage in learning from classical data \cite{Wakeham2024-uy}. However, in modern artificial intelligence research, the machine learning model is expected to adaptively discover and utilise the unknown structures in the training data with as few assumptions or biases as possible \cite{Sutton2019Bitter, Chollet2019-xi, Kaplan2020-cf}.

Despite this limitation, a growing body of work has examined quantum agents acting within classical environments \cite{Skolik2022-co, Jerbi2021-dl, Nico2025-qg, Jin2025-sg, Lazaro2025-jd, Kubo2022-gs, Kolle2024-fg}. Prior studies have explored the use of parameterised quantum circuits (PQCs) for Q-function and value-function approximation \cite{Skolik2022-co, Jin2025-sg}, as policy networks trained using REINFORCE \cite{Jerbi2021-dl}, as components of the actor and critic networks for the soft actor-critic algorithm\cite{Kolle2024-fg}, or the proximal policy optimisation (PPO) algorithm \cite{Jin2025-sg}. These works demonstrate that PQCs can, in principle, serve as function approximators within standard RL pipelines. However, they do not address a fundamental question: \emph{Does quantum entanglement improve agent performance in classical RL tasks?} Since entanglement is the resource often associated with quantum advantage, an answer to this urgent question is timely for quantum machine learning. No experimental benchmarking study has isolated and evaluated the role of entanglement in PQC-based RL agents operating in classical environments. Existing works typically introduce entanglement implicitly or treat it as a design choice rather than a variable of interest.

\subsection{Proximal Policy Optimisation}\label{sec:methods-ppo}
Our implementation of the Proximal Policy Optimisation (PPO) algorithm is directly based on CleanRL \cite{huang2022cleanrl}. The total objective of PPO that is optimised at each iteration is 
\begin{equation}
    L_t^{CLIP+VF+S} (\theta) = \mathbb{\hat{E}}_t \left[L_t^{CLIP}(\theta) - c_1 L_t^{VF} (\theta) + c_2 S[\pi_\theta](s_t)  \right],
\end{equation}
which makes use of the entropy $S[\pi_\theta](s_t)$ of the policy at state $s_t$ and the clipped surrogate objective \cite{Schulman2017-jz}
\begin{equation}
    L^{CLIP} (\theta) = \mathbb{\hat{E}}_t \left[ \min (r_t (\theta)\hat{A}_t, \text{clip}(r_t (\theta), 1-\epsilon, 1+\epsilon)\hat{A}_t )   \right],
\end{equation}
where $\theta$ is the parameters of the policy function $\pi_\theta (a_t | s_t)$, $a_t$ is the action taken at time $t$ given observed environment state $s_t$. $r_t (\theta) = \frac{\pi_\theta (a_t | s_t)}{\pi_{\theta_{\text{old}}} (a_t | s_t)} $ is the ratio function, and $c_1, c_2$ are coefficients. $\hat{A}_t$ is the estimated advantage. In \cite{Schulman2017-jz}, a truncated advantage function was used:
\begin{equation}
        \hat{A}_t = \delta_t + (\gamma \lambda)\delta_{t-1} + \cdots + \cdots + (\gamma \lambda)^{T-t+1}\delta_{T-1}.
\end{equation}
Here, $\delta_t = r_t + \gamma V(s_{t+1})- V(s_t )$, $r_t$ is the reward, $V(s)$ is a learned state-value function, trained with a squared-error loss $L_t^{VF}$ between the predicted value and the discounted cumulative reward $V_t^{\text{target}}$ associated with having taken an action $a_t$ in a given state $s_t$ and is computed based on the observations made throughout an experienced trajectory (episode), i.e. a sequence of interactions between an agent and the environment \cite{Bick2021-rx}, $\gamma$ is the temporal-difference discount factor, $T$ is the length of the episode and $\lambda$ is a hyperparameter.

\subsection{Design of the Classical-Quantum Hybrid Agent Network}

\begin{figure*}
\centering
\includegraphics[width=0.75\textwidth]{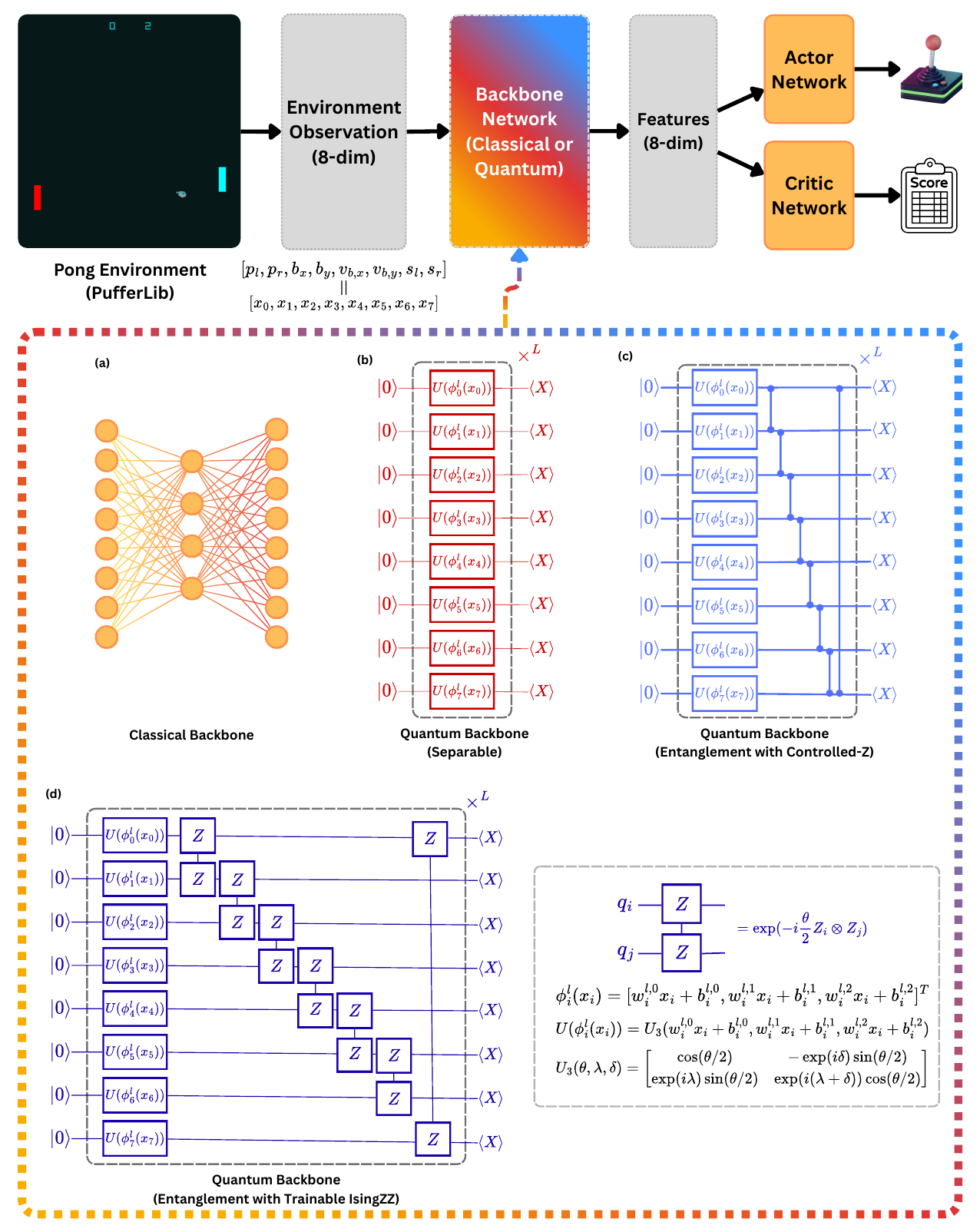}
\caption{ The overall architecture of our quantum-classical hybrid agent. The observation of the environment is an $8$-element vector $[p_l, p_r, b_x, b_y, v_{b,x}, v_{b, y}, s_l, s_r]$, where $p_l$ and $p_r$ denote the position of the left and right paddle, respectively; $b_x$ and $b_y$ are the coordinates of the ball, $v_{b,x}$ and $v_{b, y}$ are the velocity of the ball on the $x-$ and $y-$direction, respectively; $s_l$ and $s_r$ are the scores for the left and the right paddle.
A backbone network, whether classical or quantum, takes the observation vector as input and produces an 8-dimensional feature vector. A classical actor network and a classical critic network share the feature vector. Based on the features provided by the backbone network, the actor network proposes an action based on the input representations from the PQC feature extractor; the critic network provides a scalar evaluation of the current state. 
The four different kinds of backbone network structure studied in this paper are shown in (a) to (d). (a): classical multi-layer perception (MLP); (b) $8$-qubit separable parameterised quantum circuit; (c) $8$-qubit parameterised quantum circuit with fixed entanglement gates (controlled-Z gates), and (d) $8$-qubit parameterised quantum circuit with trainable entanglement gates (the IsingZZ gate $\exp{(-i \frac{\theta}{2} Z_i \otimes Z_j)}$). At the end of the parameterised quantum circuits, all qubits are measured with the Pauli X observable. 
}\label{fig:agent-arch}
\end{figure*}

Instead of using the PQC solely as a Q-function approximator or a policy function, we adopt an 8-qubit QFM circuit as a backbone, providing features for downstream actor and critic heads, following the construction of the PPO agent \cite{Schulman2017-jz} in CleanRL \cite{huang2022cleanrl}. The architecture of our agent can be found in \autoref{fig:agent-arch}. We compare the performance of our hybrid network with four qualitatively different backbone networks:
\begin{itemize}
    \item A classical multi-layer perceptron (MLP), which has three layers in total: one input, one output and one hidden.
    \item A separable data reuploading PQC, with only single-qubit gates.
    \item An entangled data reuploading PQC. The entanglement gate is the controlled-Z gate.
    \item An entangled data reuploading PQC. The entanglement gate is the trainable IsingZZ ($R_{ZZ}$) gate:
    \begin{equation}
        R_{ZZ} = e^{- i\frac{\theta}{2} Z\otimes Z}.
    \end{equation}
\end{itemize}
where each backbone takes an $8$-dimensional observation vector as input encoding the left and right paddle positions $p_l$, $p_r$, and velocities $s_l$, $s_r$, as well as the ball position $b_x$, $b_y$ and velocity $v_{b,x}$, $v_{b,y}$. The output is a new $8$-dimensional feature vector that represents the features extracted from the game state and which is passed to the classical actor-critic framework for policy implementation and scoring.

Each of these $4$ different types of backbones can be configured to have a different number of trainable parameters. For the classical MLP backbone, the number of total trainable parameters is controlled by the number of neurons in the hidden layer (the dimension of the hidden-layer representation), since the number of neurons in the input and output layers is both $8$. For the parameterised quantum circuits, the number of parameters can be controlled by changing the number of layers. For all three types of PQCs, the majority of trainable parameters are encoded in single-qubit gates that can also take input data as arguments. Each element of the observation vector $\mathbf{x}$ is first mapped to three values with six trainable parameters $w^{l, 0}_i, w^{l, 1}_i, w^{l, 2}_i, b^{l,0}_i, b^{l,1}_i, \text{ and } b^{l,2}_i$:
\begin{equation}
    \phi_i^l (x_i) = [w^{l, 0}_i x_i + b^{l,0}_i, w^{l, 1}_i x_i + b^{l,1}_i, w^{l, 2}_i x_i + b^{l,2}_i ]^T,
\end{equation}
where $i$ is the index of the observation vector element, as well as the index of the qubit, and $l$ denotes the layer number. Then, each element of the $\phi_i^l (x_i)$ vector is used as inputs to the $U_3$ gate:
\begin{equation}
    U (\phi_i^l (x_i)) = U_3 (w^{l, 0}_i x_i + b^{l,0}_i, w^{l, 1}_i x_i + b^{l,1}_i, w^{l, 2}_i x_i + b^{l,2}_i),
\end{equation}
where 
\begin{equation}
U_3 (\theta, \lambda, \delta)=
    \begin{bmatrix}
        \cos \frac{\theta}{2} & -\exp(i\delta) \sin(\frac{\theta}{2}) \\
        \exp(i \lambda)\sin(\frac{\theta}{2}) & \exp(i (\lambda + \delta))\cos(\frac{\theta}{2})
    \end{bmatrix}.
\end{equation}
\newline
Hence, for separable and CZ-entangled PQCs, each layer has $6\times 8 = 48$ trainable parameters. For IsingZZ-entangled PQCs, each layer has $6\times 8 + 8 = 56$ trainable parameters.

\section{Results}\label{sec:results}

We systematically evaluated the effect of quantum entanglement on the performance of a quantum–classical hybrid reinforcement learning agent in the Pong environment. To isolate the role of entanglement, we compared three classes of quantum backbones (separable, CZ-entangled, and IsingZZ-entangled) against classical MLP baselines across a wide range of parameter counts.

All reported results are averaged over 10 independent training runs with different random initialisations. Performance is evaluated using the episodic return at the final stage of training, where the maximum achievable return in Pong is $+21$, and the minimum is $-21$. A summary of all configurations and their final performance statistics is provided in \autoref{tab:performance-res}.

\begin{table*}
\caption{Averaged Performance for our hybrid agent with different parameter configurations. For each configuration, the statistics (mean and standard deviation) of agent performance are calculated from 10 random initialisations of the trainable parameters. For the Pong game, the maximum achievable return is 21, and the minimum is -21. And for single runs before smoothing with an exponential moving average, the return values are always integers. For quantum backbones, the number of parameters is shown as $\textit{number of parameters in a single layer}\times\textit{number of layers}$. For classical backbones, since they all have three layers, the number of parameters is shown as $\textit{input layer dimension}\times\textit{hidden layer dimension}+\textit{hidden layer dimension}\times\textit{output layer dimension}$. The best performance of all different configurations comes from the classical backbone with $4096$ parameters, which is supposed to have access to the same ``hidden dimension size" as an 8-qubit quantum circuit, which also has a $2^8$-dimensional state vector. For all other configurations besides the classical backbone with $4096$ parameters, the minimum reward at the final step during the training process is $-21$, while the classical $4096$-parameter MLP backbone has a minimum final reward of $-5$ across $10$ different random initialisations. The data in the ``Final Return Max." column is from the raw training statistics before being smoothed with a weighted exponential moving average, as those in \autoref{fig:quant-backbone-compare} and \autoref{fig:quant-vs-mlp-supp}.
}\label{tab:performance-res}%
\begin{ruledtabular}
\begin{tabular}{ccccc}

{Backbone Type} & {Param. Num.}  &{Final Return Mean} & {Final Return Std.} & {Final Return Max.}\\ \hline

Separable & $48 = 48\times 1$ & -17.9 & 4.72 &  -8\\
Separable & $96 = 48\times 2$ & -17.9 & 4.68  & -8 \\
Separable & $144 = 48\times 3$ & -19.5 & 4.50 & -6\\
Separable & $192=48\times 4$ & -21.0 & 0.00 & -21\\
Separable & $240 = 48\times 5$ & -20.4 & 1.80 & -15\\
Separable & $288 = 48\times 6$ & -20.8 & 0.60 & -19\\

CZ-entangled & $48 = 48\times 1$ & -10.0 & 13.39 & 13\\
CZ-entangled & $96 = 48\times 2$ & -2.1 & 15.02 & 17\\
CZ-entangled & $144 = 48\times 3$ & -1.0 & 12.61 & 17\\
CZ-entangled & $192=48\times 4$ & -10.0 & 12.07 & 16\\
CZ-entangled & $240 = 48\times 5$ & -14.8 & 4.96  & -3\\
CZ-entangled & $288 = 48\times 6$ & -15.7 & 6.29  & -2\\

IsingZZ-entangled & $56 = 56\times 1$ & -5.0 & 12.63  & 12\\
IsingZZ-entangled & $112 = 56\times 2$ & -9.7 & 10.91 & 9\\
IsingZZ-entangled & $ 168 = 56\times 3$ & -10.2 & 9.40 & 4\\
IsingZZ-entangled & $ 224= 56 \times 4$ & -14.4 & 8.34 & 2\\
IsingZZ-entangled & $ 280 = 56\times 5$ & -13.4 & 11.83 & 19\\
IsingZZ-entangled & $ 336 = 56\times 6$ & -16.8 & 6.71 & -1\\

Classical    & $64=8\times 4 + 4\times 8$   & -8.7  &11.79 & 8  \\
Classical   & $128 = 8\times 8 + 8\times 8$   & -6.7  & 9.41 & 11 \\
Classical   & $256 = 8\times 16 + 16\times 8$   & -3.0  & 9.69 & 12 \\
Classical   & $336 = 8\times 21+21\times 8$   & -0.8  & 11.14 & 14\\
Classical   & $4096 = 8 \times 2^8 + 2^8 \times 8$  & 15.0  & 7.03 & 19\\


\end{tabular}
\end{ruledtabular}
\end{table*}

\subsection{Entangled PQC backbones consistently outperform separable circuits}

\begin{figure*}
    \centering
    \begin{subfigure}{0.3\textwidth}
        \includegraphics[width=\textwidth]{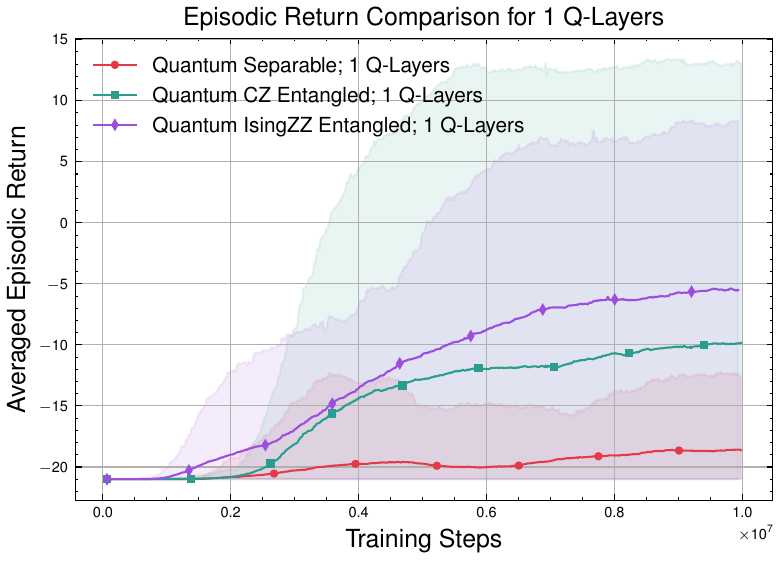}
        \caption{ }
    \end{subfigure}
    \begin{subfigure}{0.3\textwidth}
        \includegraphics[width=\textwidth]{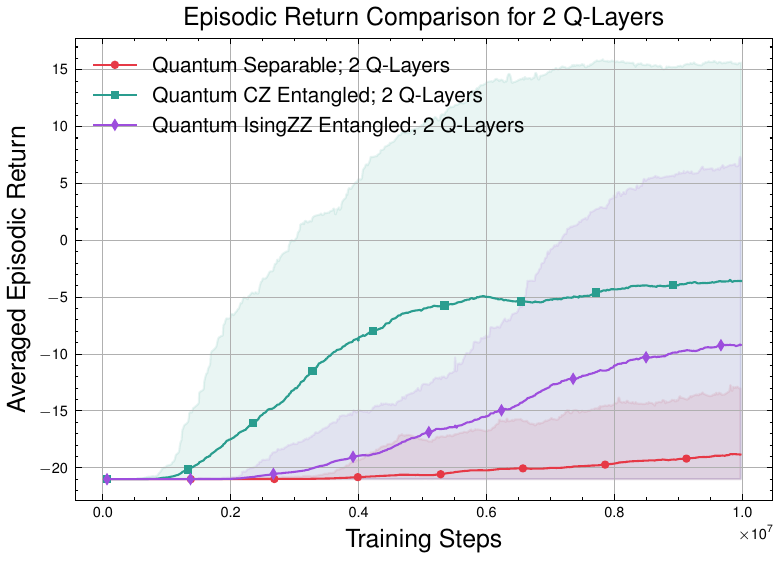}
        \caption{ }
    \end{subfigure}
    \begin{subfigure}{0.3\textwidth}
        \includegraphics[width=\textwidth]{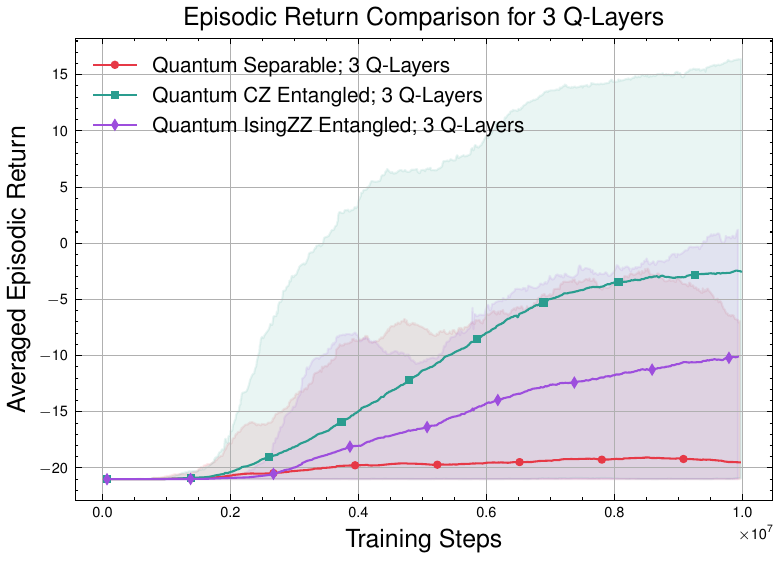}
        \caption{ }
    \end{subfigure}
    \hfill
    \begin{subfigure}{0.3\textwidth}
        \includegraphics[width=\textwidth]{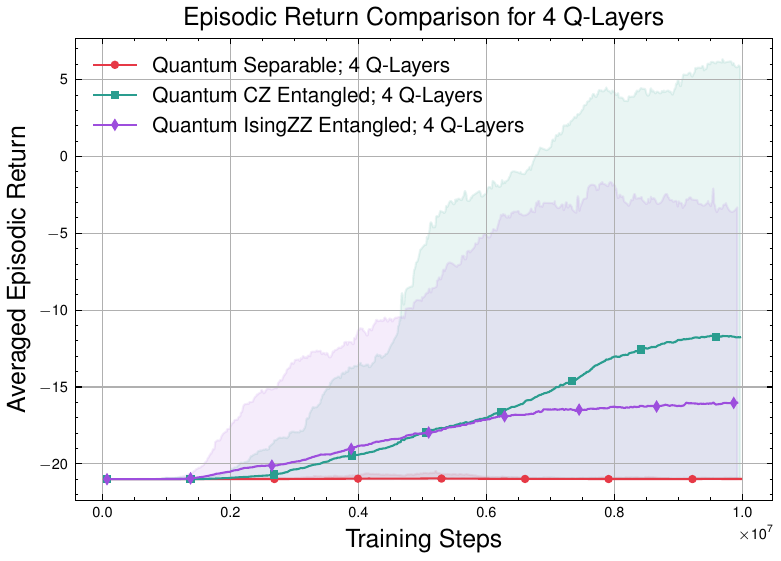}
        \caption{ }
    \end{subfigure}
    \begin{subfigure}{0.3\textwidth}
        \includegraphics[width=\textwidth]{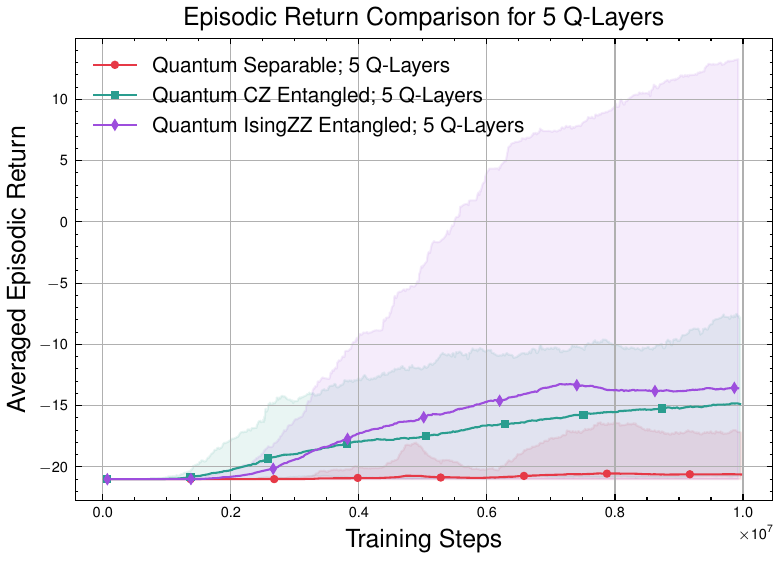}
        \caption{ }
    \end{subfigure}
    \begin{subfigure}{0.3\textwidth}
        \includegraphics[width=\textwidth]{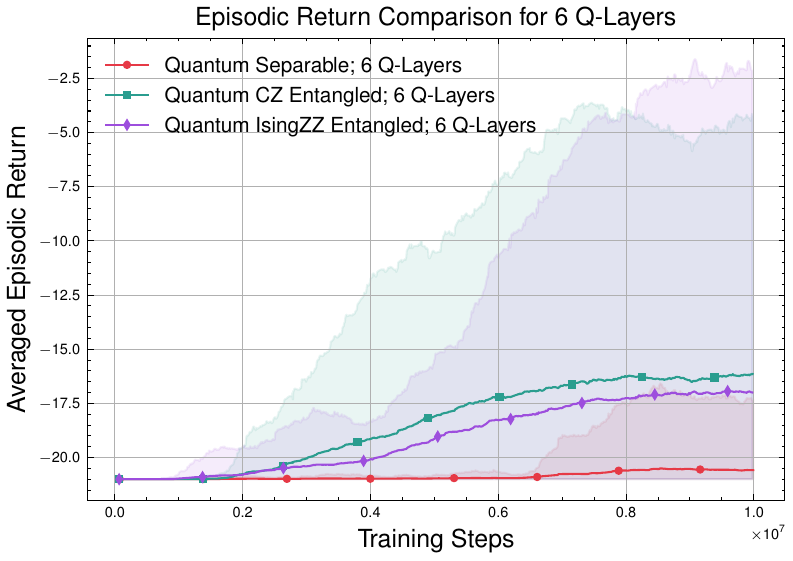}
        \caption{ }
    \end{subfigure}
    \caption{
    The return of different quantum backbone configurations with respect to the total training steps. The curves for separable, CZ-entangled, and IsingZZ-entangled are averaged over 10 runs with different random initialisations. The averaged return curve and the max-min shade for returns are calculated and plotted from the smoothed raw data using a weighted exponential moving average. We can see that, although entangled quantum backbones could, on average, outperform separable ones, it is unclear, based on the current results, whether trainable entanglement gates (IsingZZ) could achieve better performance than fixed entanglement gates (CZ). Even though, on average, the backbone with IsingZZ gates can outperform that with CZ gates (as in (a)), the best performance of quantum circuits with CZ gates is better than that of backbones with IsingZZ gates.
    }
    \label{fig:quant-backbone-compare}
\end{figure*}

Across all tested configurations, quantum backbones with entanglement substantially outperform separable PQCs with comparable numbers of parameters. This trend is evident in both the averaged learning curves (\autoref{fig:quant-backbone-compare}) and the final episodic returns reported in \autoref{tab:performance-res}. The statistics of the episodic length can be found in Appendix A (\autoref{tab:episodic-length-res}).

Separable PQC backbones exhibit consistently poor performance regardless of circuit depth. As shown in \autoref{fig:quant-backbone-compare}(a–f), increasing the number of layers does not lead to meaningful improvements in averaged episodic return; in several cases, performance collapses to the minimum achievable return of $-21$. This behaviour is also reflected in \autoref{tab:performance-res}, where separable circuits achieve the lowest mean and maximum final returns across all parameter counts.

In contrast, both CZ-entangled and IsingZZ-entangled backbones achieve significantly higher returns. For the same number of layers, entangled PQCs learn faster and converge to better-performing policies, as demonstrated by the separation between entangled and separable learning curves in \autoref{fig:quant-backbone-compare}. Importantly, this improvement cannot be attributed to differences in single-qubit parameterisation as separable and entangled PQCs share identical gate topologies that differ only in the absence and presence (respectively) of two-qubit entangling operations. This strongly suggests that entanglement is responsible for the observed performance gains. Without entanglement, the representations generated by the quantum circuit are restricted to element-wise nonlinear transformations of the input vector. In the entangled case, however, the representation is more flexible and can be extended by non-linear combinations of input elements.

\subsection{Optimal performance occurs at shallow circuit depths}

For entangled PQC backbones, increasing circuit depth does not monotonically improve performance. Instead, both averaged and maximum episodic returns peak at relatively shallow depths. For CZ-entangled backbones, the best overall performance is achieved with two to three layers, while deeper circuits lead to reduced average returns and higher variance across random initialisations. A similar pattern is observed for IsingZZ-entangled backbones, where the highest mean performance occurs at a single layer, although the maximum return across runs peaks at deeper layers.

These results suggest that deeper quantum circuits may suffer from optimisation difficulties, consistent with the barren plateau phenomenon \cite{McClean2018-pf}, which can severely degrade gradient-based training. Reinforcement learning appears particularly sensitive to this effect, as the training signal is noisier and less structured than in supervised learning tasks.

Importantly, the presence of trainable entanglement gates (IsingZZ) does not guarantee superior performance compared to fixed entanglement (CZ). While IsingZZ backbones occasionally achieve higher peak returns, their average performance is comparable to or slightly worse than CZ-entangled circuits, indicating a trade-off between expressivity and trainability.

\subsection{Entangled PQCs can outperform classical baselines in the low-parameter regime}

\begin{figure*}
    \centering
    \begin{subfigure}{0.3\textwidth}
        \includegraphics[width=\textwidth]{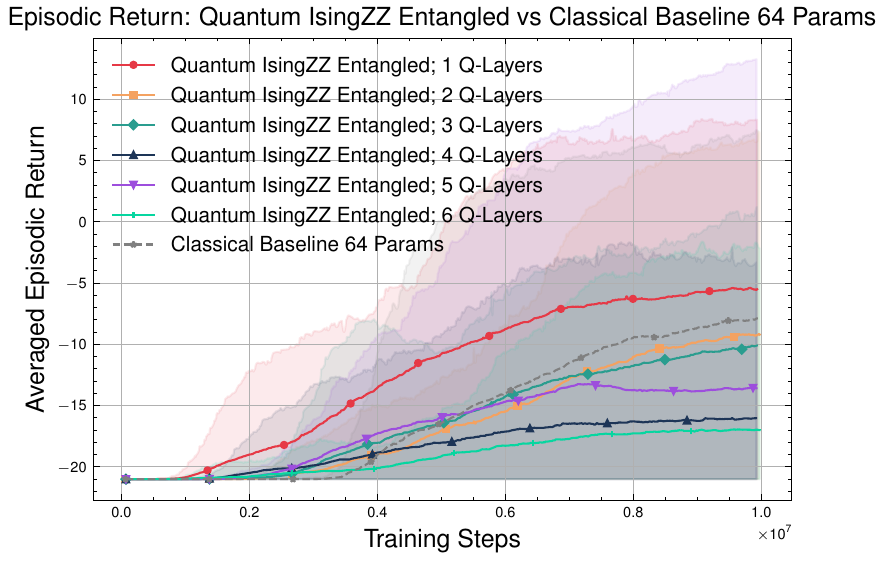}
        \caption{ }
    \end{subfigure}
    \begin{subfigure}{0.3\textwidth}
        \includegraphics[width=\textwidth]{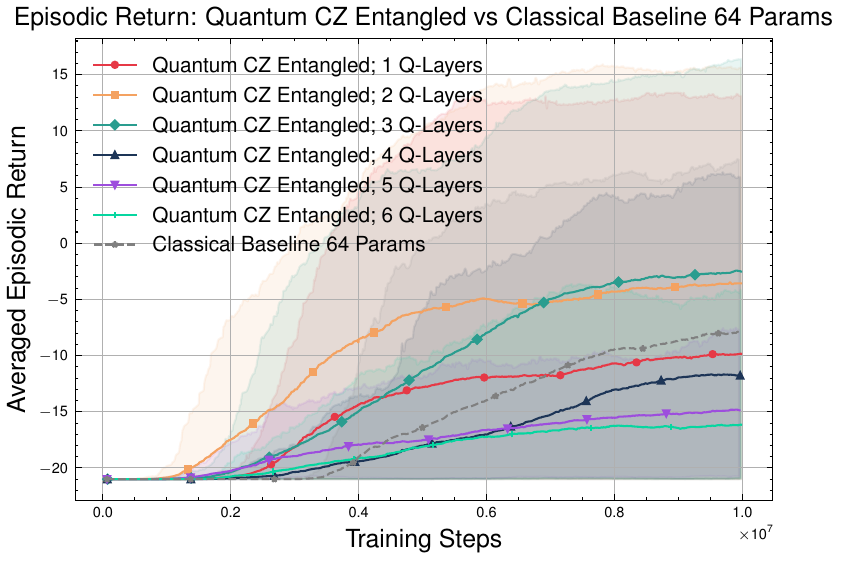}
        \caption{ }
    \end{subfigure}
    \begin{subfigure}{0.3\textwidth}
        \includegraphics[width=\textwidth]{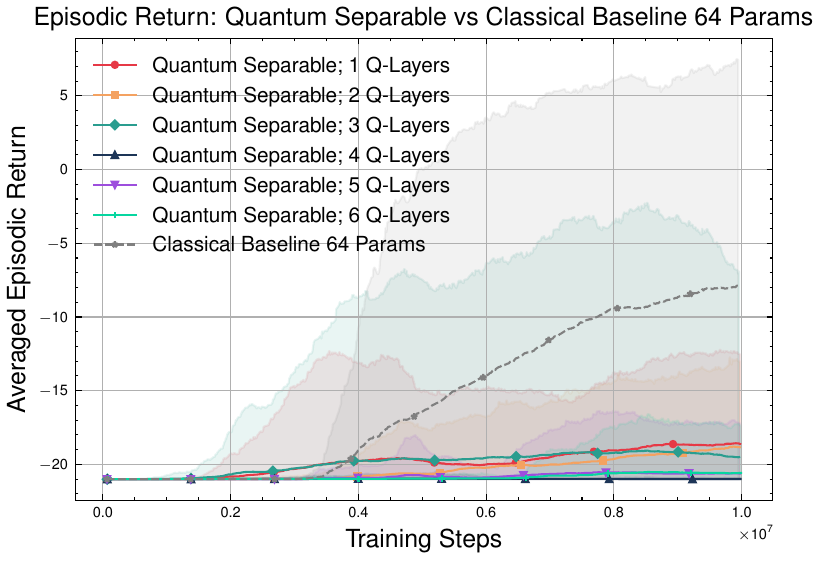}
        \caption{ }
    \end{subfigure}
    \caption{
        Comparing the (averaged) episodic performance of the three different quantum backbones (separable, CZ-entangled and IsingZZ-entangled) w.r.t. classical MLP with $64$ parameters. {(a):} The performance of the IsingZZ-entangled backbone, compared with the classical MLP backbone with $64$ parameters. Each layer of the IsingZZ-entangled backbone has $56$ parameters. In this configuration, only the $1$-layer one could outperform the classical baseline, but with fewer parameters ($56$ vs $64$). In this case, increasing the number of layers (and hence the number of parameters) does not guarantee improved performance either. {(b):} The performance of the CZ-entangled backbone, compared with the classical MLP backbone with $64$ parameters. Among the different layer configurations, those with $2$ and $3$ layers can outperform the classical baseline after averaging results across $10$ random initialisations. Since CZ-entangled quantum circuits have $48$ parameters per layer, the CZ-entangled backbones that outperform the classical baseline have $96$ and $144$ parameters, respectively. However, when the number of layers is increased beyond $3$, the performance of the quantum-classical hybrid agent drops by half on average at the final iteration. {(c):} The performance of the separable backbone, compared with the classical MLP backbone. We can see that there is little difference in performance across different layer configurations of the separable backbone, and they all have worse average returns than the $64$-parameter MLP backbone.
    }
    \label{fig:quant-vs-mlp}
\end{figure*}

In the low-parameter regime, entangled quantum backbones can match or exceed the performance of classical MLP baselines, despite having comparable or even fewer trainable parameters (\autoref{fig:quant-vs-mlp}). Specifically, a 1-layer IsingZZ-entangled PQC with 56 parameters outperforms a classical MLP with 64 parameters (\autoref{fig:quant-vs-mlp}(a)); $2$- and $3$-layer CZ-entangled PQCs outperform the same classical baseline after averaging across random initialisations, albeit with slightly higher parameter counts (\autoref{fig:quant-vs-mlp}(b)).

In contrast, separable PQCs fail to outperform the classical baseline in any tested configuration, further reinforcing the central role of entanglement. These results demonstrate that quantum-enhanced feature extraction can provide tangible benefits when model capacity is constrained, a regime that is particularly relevant for near-term quantum hardware.

At higher parameter counts, classical networks ultimately achieve superior absolute performance. A large classical MLP with $4096$ parameters consistently outperforms all quantum backbones, reflecting the greater flexibility and optimisation stability of classical models at scale (\autoref{fig:quant-vs-mlp-supp}).

\subsection{Quantum and classical backbones learn fundamentally different representations}

\begin{figure*}
\centering
\includegraphics[width=0.7\textwidth]{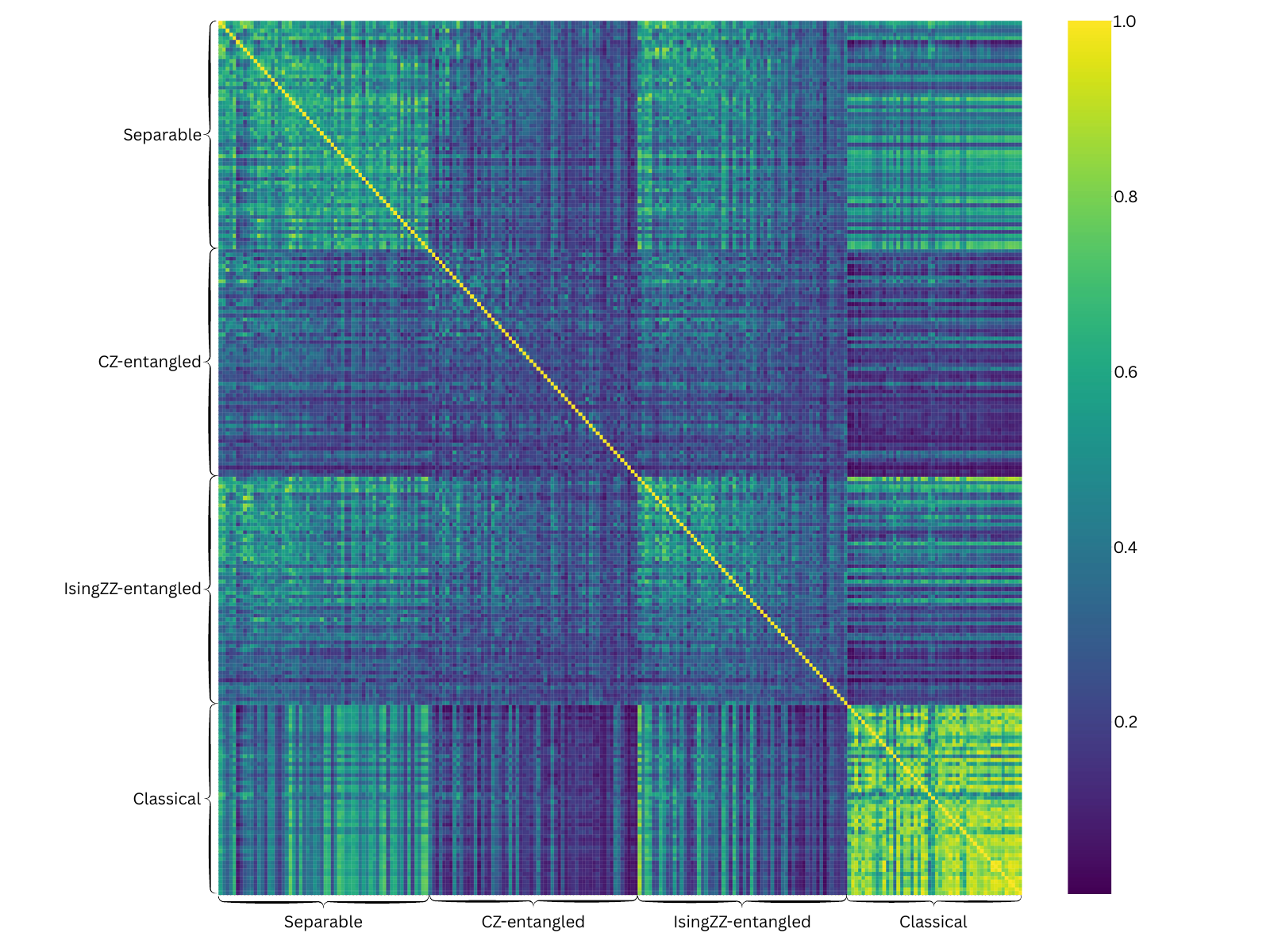}
\caption{ The similarities of representations generated by different backbone configurations, calculated with centred kernel alignment (CKA) \cite{kornblith2019similarityneuralnetworkrepresentations}. 
The heatmap is divided into approximately sixteen blocks, representing intragroup (diagonal blocks) and intergroup (off-diagonal blocks) similarities.
For intra-group similarities, we see that the classical backbone generally produces similar representations, even though the hidden dimension differs. 
The CZ-entangled backbones produce representations with the most variability between different initialisations (intragroup block) and different backbone topologies (intergroup).
Generally, for classical neural networks, CKA would yield very high similarity scores for representations produced by networks with the same structure, trained on the same dataset, but with different random initialisations, as shown in the bottom-right corner of the heatmap. However, such intra-group similarity is much harder to find among the representations produced by the quantum backbones. Separable and IsingZZ-entangled PQC backbones have an overall slightly higher intra-group similarity score compared to the CZ-entangled ones. However, all three types of PQC-based backbone networks produce representations that share little similarity with those from the classical backbone networks.
}\label{fig:cka-heatmap}
\end{figure*}

To better understand the origin of the observed performance differences, we analysed the similarity of learned representations using centred kernel alignment (CKA) \cite{kornblith2019similarityneuralnetworkrepresentations}, as shown in the heatmap in \autoref{fig:cka-heatmap}. 

Let $X\in \mathbb{R}^{n\times p_1}$ denote representations with dimension $p_1$ for $n$ inputs, and let $Y\in \mathbb{R}^{n\times p_2}$ denote representations with dimension $p_2$ for the same $n$ inputs. The similarity between two different representations generated by two different neural networks for the same set of inputs can be measured with the linear centred kernel alignment (linear CKA). Following \cite{kornblith2019similarityneuralnetworkrepresentations}, we can calculate linear CKA scores after centreing $X, Y$ to $\hat{X}, \hat{Y}$:
\begin{equation}
    \text{CKA}_{\text{Linear}} = \frac{|| A ||_F^2}{|| B ||_F^2 || C ||_F^2},
\end{equation}
where $A = \hat{Y}^T\hat{X}, B=\hat{X}^T\hat{X}, C=\hat{Y}^T\hat{Y}$, and $||\cdot ||_F$ denotes the Frobenius norm for matrices.

Classical MLP backbones produce highly similar representations across different random initialisations and hidden dimensions, consistent with prior observations in deep learning. In contrast, representations generated by quantum backbones exhibit low intra-group similarity, particularly for CZ-entangled circuits, indicating a high degree of representational diversity.

Moreover, all quantum backbones—separable and entangled—produce representations that are markedly dissimilar from those learned by classical networks. This suggests that PQCs do not simply approximate classical solutions, but instead explore qualitatively different regions of representation space.

Among quantum architectures, separable PQCs yield representations that are more similar to classical ones than those produced by entangled circuits. Entangled PQCs, especially with CZ gates, generate the most distinct representations, aligning with their superior task performance.

\section{Discussion}\label{sec:discussion}

In this work, we presented a controlled experimental study of the role of quantum entanglement in quantum–classical hybrid reinforcement learning agents operating in a classical environment. By explicitly isolating entanglement as a design variable in parameterised quantum circuit (PQC) backbones, we demonstrated that entanglement is not merely an architectural embellishment, but a functional computational resource that can improve agent performance. Our results show that PQC backbones with entanglement consistently outperform separable PQCs with comparable parameter counts when learning to play Pong. This finding provides direct empirical evidence that entanglement enhances feature extraction for reinforcement learning tasks, addressing a central and previously unresolved question in quantum reinforcement learning research.

\subsection{Why entanglement helps in classical reinforcement learning}
The Pong environment used in this study provides an 8-dimensional observation vector, where each component encodes a distinct aspect of the game state (paddle positions and velocities, ball position, and ball velocity). Although these variables are presented independently, optimal decision making depends critically on their joint relationships—for example, how paddle motion should respond to the ball’s trajectory.

In separable PQC backbones, each observation component is processed independently by a corresponding qubit, and the final representation is obtained through single-qubit measurements. In this setting, there is no mechanism for modelling interactions between different state variables during feature extraction. As a result, separable circuits struggle to construct useful representations, regardless of circuit depth.

Entangling gates fundamentally change this behaviour. Both CZ and IsingZZ entanglement introduce non-classical correlations between qubits, enabling the PQC to encode joint information about multiple input variables. From a functional perspective, entanglement acts as an interaction mechanism analogous to multiplicative feature coupling in classical neural networks. This provides a plausible explanation for the substantial performance gap observed between separable and entangled PQC backbones.

\subsection{Representation learning beyond classical solutions}
Our representation similarity analysis further supports the interpretation that entangled PQCs learn qualitatively different representations from classical neural networks. While classical MLP backbones converge to highly similar representations across different initialisations and hidden dimensions, quantum backbones, particularly those with entanglement, exhibit low intra-group similarity and minimal overlap with classical representations.

This observation suggests that entangled PQCs do not simply approximate classical feature extractors with fewer parameters. Instead, they explore distinct regions of representation space that are not easily accessible to classical architectures of comparable size. Importantly, the PQC representations that are most dissimilar from classical ones, those produced by CZ-entangled circuits, are also those that yield the strongest performance improvements.

These findings align with recent work on the expressive properties of quantum models, which argues that entanglement enables access to better function spaces
even when the number of trainable parameters is limited. Our results extend this perspective to reinforcement learning, where representation quality is critical for downstream policy optimisation.

\subsection{Trade-offs between expressivity and trainability}
Although entanglement improves performance, our results also highlight important limitations. Increasing circuit depth does not reliably improve learning outcomes, and deeper entangled PQCs often perform worse than their shallower counterparts. This behaviour is consistent with the barren plateau phenomenon, which causes gradients to vanish as circuit depth increases and severely hinders optimisation.

Interestingly, introducing trainable entanglement through IsingZZ gates does not consistently outperform fixed CZ entanglement. While IsingZZ-entangled circuits occasionally achieve higher peak returns, their average performance is comparable to or worse than CZ-based circuits. This suggests a trade-off between expressivity and trainability: additional trainable parameters increase model flexibility but also exacerbate optimisation difficulties.

These results indicate that, for reinforcement learning tasks, shallow entangled circuits may represent a sweet spot, offering sufficient expressivity without rendering training intractable.

\subsection{Quantum advantage in the low-parameter regime}

One of the most practically relevant findings of this study is that entangled PQC backbones can outperform classical MLP baselines in the low-parameter regime. In particular, we observe configurations where entangled PQCs achieve higher average returns than classical networks with similar or even fewer parameters.

This result does not constitute a full quantum advantage in the complexity-theoretic sense. However, it does demonstrate a resource advantage: entanglement enables more effective use of limited parameter budgets. Such regimes are precisely those relevant to near-term quantum hardware, where circuit depth, coherence time, and parameter counts are tightly constrained.

At larger parameter counts, classical networks retain a decisive advantage. A sufficiently large MLP consistently outperforms all quantum backbones tested in this study, reflecting the greater expressive freedom and optimisation stability of classical models. A recent analysis of the Fourier fingerprints of the Quantum Fourier Model (QFM) \cite{Strobl2025-ea} shows that, due to trainability constraints, QFM circuits admit only a polynomial scaling of independent trainable parameters with the number of qubits. As a consequence, correlations arise among the model’s Fourier coefficients, effectively restricting the number of frequencies that can be independently controlled by the parameters, despite the fact that the total number of accessible frequencies grows exponentially with the number of qubits. Classical models do not have such limitations, giving them more degrees of freedom. This result is neither surprising nor discouraging; rather, it underscores the importance of identifying application domains and operating regimes where quantum resources offer complementary benefits rather than wholesale replacement of classical methods.

\subsection{Scope, limitations, and future directions}
This study is intentionally narrow in scope. We focused on a low-dimensional classical environment to tightly control experimental variables and enable systematic benchmarking of entanglement. More complex environments, particularly those involving high-dimensional observations such as images or videos, would require either quantum models capable of direct high-dimensional encoding or hybrid pipelines that rely heavily on classical preprocessing.

To better reflect the competitive nature of Pong, an important next step is to move beyond training and evaluating against a single fixed opponent. For example, the agent could be trained and tested against a range of opponents, such as self-play (playing against a copy of itself) or an opponent pool consisting of multiple policies (e.g., different random seeds or saved checkpoints). Evaluating performance across this opponent set would help distinguish genuine improvements in learning from overfitting to one particular opponent behaviour, and would provide a more robust test of whether entanglement improves performance in competitive settings.

Scaling PQC-based reinforcement learning to such settings remains an open challenge. Current quantum simulators and hardware are not yet capable of efficiently training large-scale quantum models with realistic data volumes. Moreover, optimisation challenges such as barren plateaus become increasingly severe as circuit size grows. These observations motivate several directions for future investigation, including:
\begin{itemize}
    \item Alternative entanglement patterns and circuit topologies that improve trainability,
    \item Task domains where low-parameter efficiency is especially valuable,
    \item Multi-agent/self-play competitive evaluation, where both players learn (self-play or population-based training) and performance is reported not only as mean episodic return but also robustness and stability under opponent adaptation (Markov-game perspective),
    \item Training strategies that mitigate barren plateaus in reinforcement learning settings,
    \item Other hybrid architectures that combine quantum feature extractors with classical representation learning.
\end{itemize}

Overall, our work demonstrates that quantum entanglement can meaningfully enhance reinforcement learning performance in classical environments, not by outperforming large classical models outright, but by enabling more efficient representation learning under constrained resources. Our findings position entanglement as a tangible and measurable contributor to quantum-enhanced decision making, and provide a concrete experimental foundation for future research in quantum reinforcement learning.

\appendix

\section{Additional Table and Figures for the RL Training Process}\label{sec:appendixA}

\begin{table*}
\caption{Averaged episodic length for our hybrid agent with different parameter configurations. For each configuration, the statistics (mean and standard deviation) of the lengths of episodes are calculated from 10 random initialisations of the trainable parameters. And for single runs before smoothing with an exponential moving average, the number of steps in a single episode is always an integer. It should be noted that the episodic length is a less informative indicator for the agents' performance compared to the actual return the agent obtains from the environment. When the number of steps in an episode is low, the agent could perform well or poorly, since in Pong, the agent needs to play with a computer-controlled paddle to win the game. Small episodic length could indicate that the agent finishes the opponent quickly, or gets finished by the opponent quickly.
}\label{tab:episodic-length-res}%
\begin{ruledtabular}
\begin{tabular}{cccccc}

{Backbone Type} & {Params }  &{Episodic Length Mean} & {Episodic Length Std.} & {Episodic Length Max.} & {Episodic Length Min.}\\ \hline

Separable & $48$ & 274.5 & 106.21 & 417 & 132 \\
Separable & $96 $ & 331.5 & 115.96  & 607 & 227 \\
Separable & $144$ & 274.5 & 114.40 & 417 & 132 \\
Separable & $192$ & 227.0 & 73.59 & 417 & 132 \\
Separable & $240$ & 236.5 & 78.91 & 417 & 132 \\
Separable & $288$ & 303.0 & 254.20 & 987 & 132 \\

CZ-entangled & $48$ & 364.6 & 149.28 & 654 & 132\\
CZ-entangled & $96$ & 431.0 & 170.13 & 892 & 274\\
CZ-entangled & $144$ & 407.3 & 161.46 & 702 & 132\\
CZ-entangled & $192$ & 378.9 & 174.92 & 797 & 132\\
CZ-entangled & $240$ & 322.0 & 127.46  & 607 & 132\\
CZ-entangled & $288$ & 616.5 & 744.95  & 2792 & 132\\

IsingZZ-entangled & $56$ & 369.3 & 168.50  & 607 & 132\\
IsingZZ-entangled & $112$ & 317.1 & 179.60 & 702 & 132\\
IsingZZ-entangled & $ 168$ & 502.4 & 206.78 & 892 & 227\\
IsingZZ-entangled & $ 224$ & 697.2 & 628.90 & 2507 & 227\\
IsingZZ-entangled & $ 280$ & 649.7 & 560.60 & 1937 & 227 \\
IsingZZ-entangled & $ 336$ & 521.5 & 508.85 & 1937 & 227\\

Classical    & $64$   & 298.1  & 172.70 & 559 & 132 \\
Classical   & $128$   & 393.1  & 244.13 & 1034 & 132 \\
Classical   & $256$   & 421.5  & 173.21 & 892 & 227\\
Classical   & $336$   & 473.8  & 280.79 & 1224 & 227\\
Classical   & $4096$  & 544.8  & 190.02 & 939 & 274 \\


\end{tabular}
\end{ruledtabular}
\end{table*}

\begin{figure*}[h!]
    \centering
    \begin{subfigure}{0.31\textwidth}
        \includegraphics[width=\textwidth]{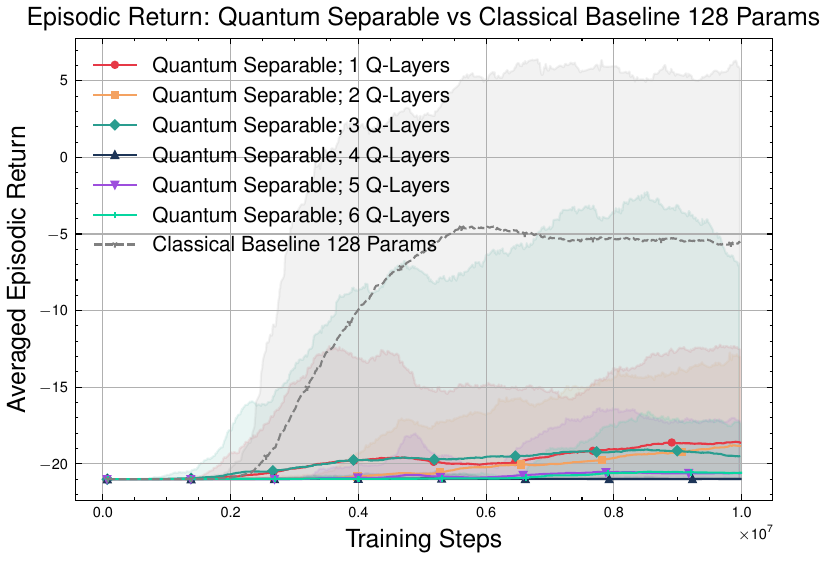}
        \caption{ }
    \end{subfigure}
    \hfill
    \begin{subfigure}{0.32\textwidth}
        \includegraphics[width=\textwidth]{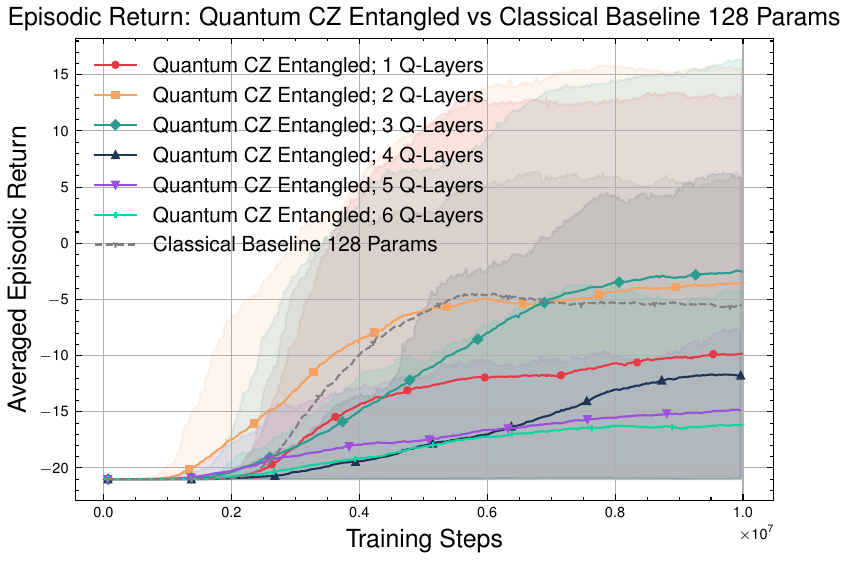}
        \caption{ }
    \end{subfigure}
    \hfill
    \begin{subfigure}{0.33\textwidth}
        \includegraphics[width=\textwidth]{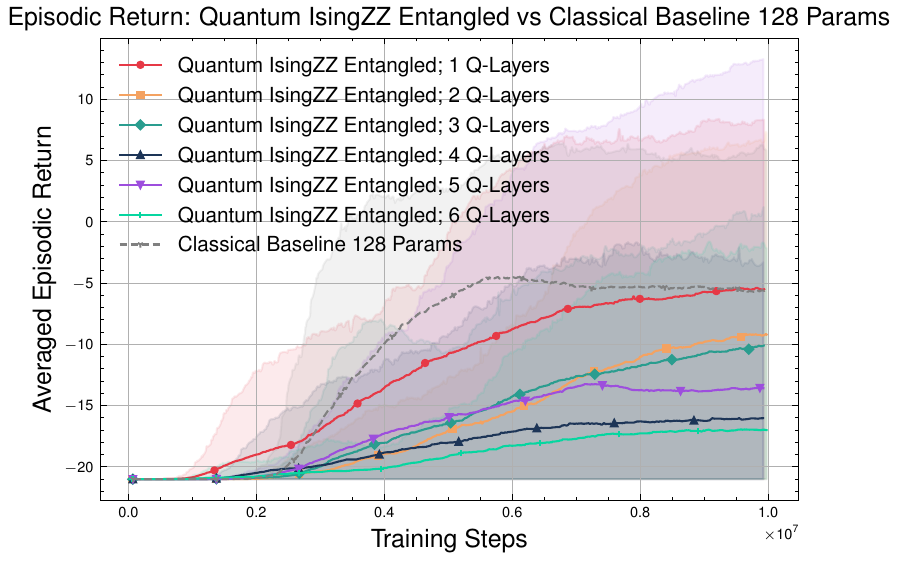}
        \caption{ }
    \end{subfigure}
    \vspace{0.2em}

    \begin{subfigure}{0.31\textwidth}
        \includegraphics[width=\textwidth]{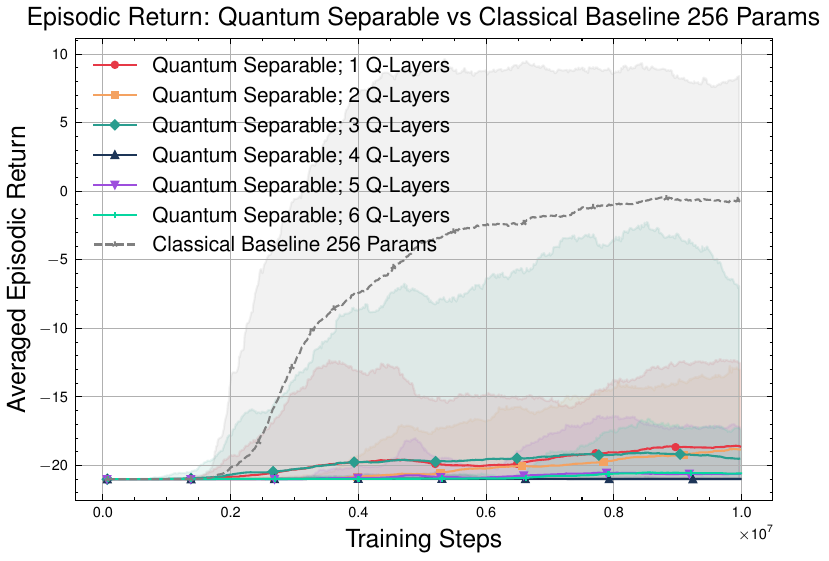}
        \caption{ }
    \end{subfigure}
    \hfill
    \begin{subfigure}{0.32\textwidth}
        \includegraphics[width=\textwidth]{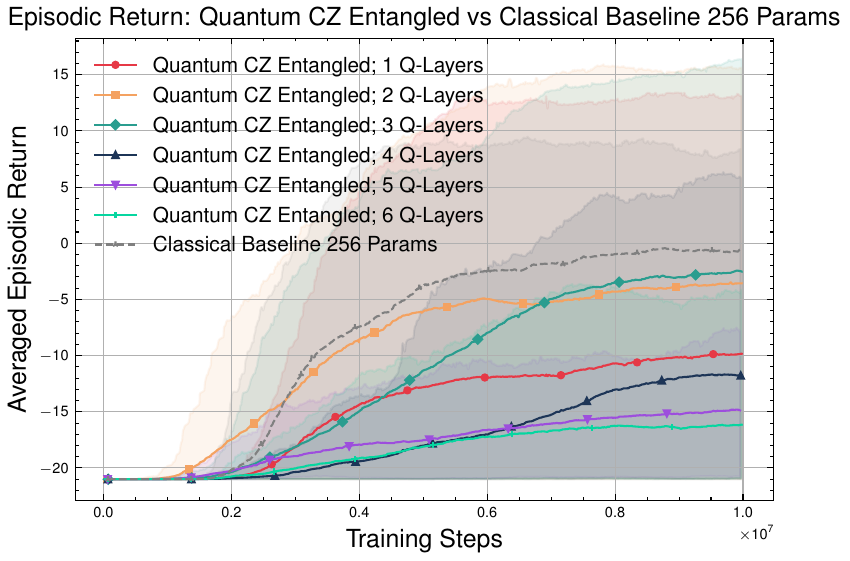}
        \caption{ }
    \end{subfigure}
    \hfill
    \begin{subfigure}{0.33\textwidth}
        \includegraphics[width=\textwidth]{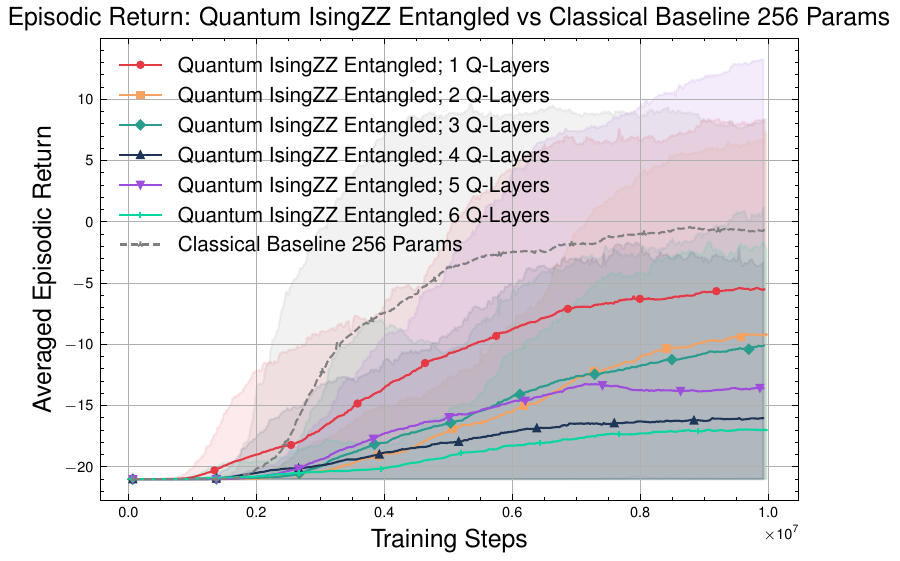}
        \caption{ }
    \end{subfigure}
    \vspace{0.2em}

    \begin{subfigure}{0.31\textwidth}
        \includegraphics[width=\textwidth]{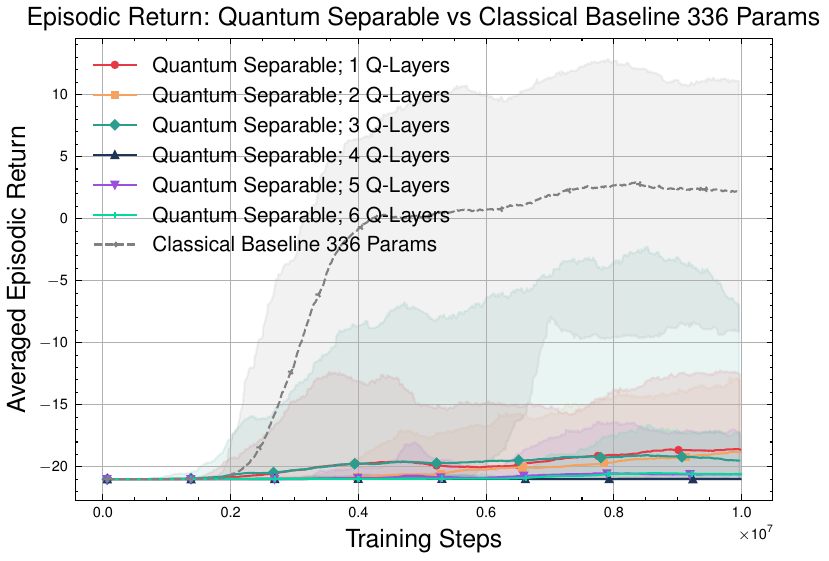}
        \caption{ }
    \end{subfigure}
    \hfill
    \begin{subfigure}{0.32\textwidth}
        \includegraphics[width=\textwidth]{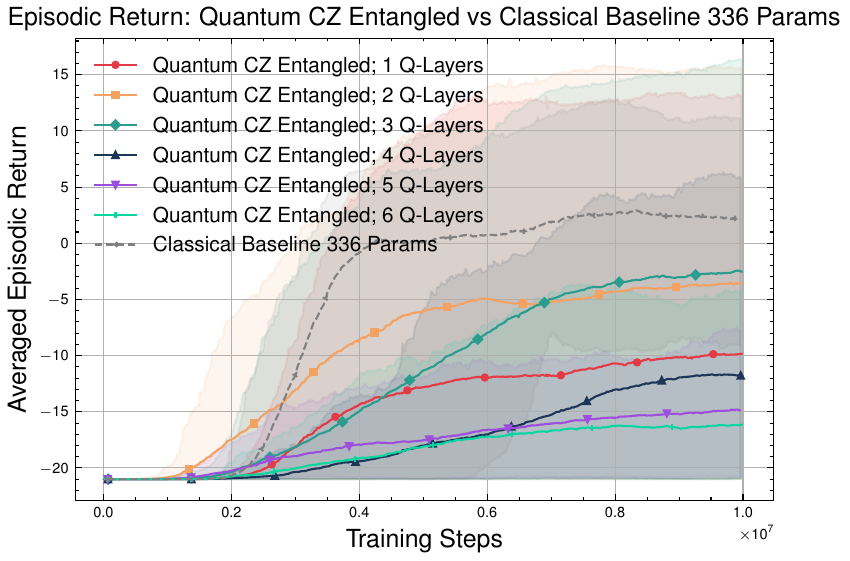}
        \caption{ }
    \end{subfigure}
    \hfill
    \begin{subfigure}{0.33\textwidth}
        \includegraphics[width=\textwidth]{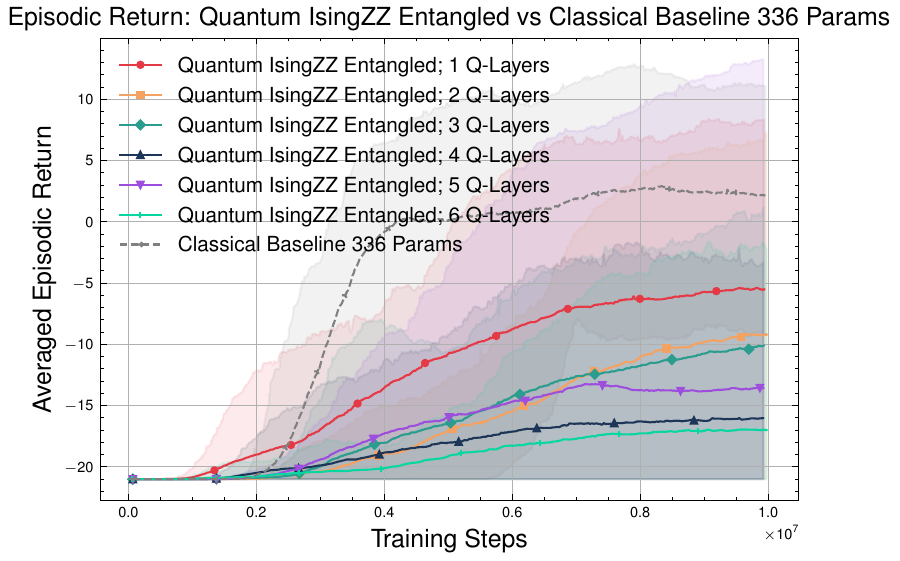}
        \caption{ }
    \end{subfigure}
    \vspace{0.2em}

    \begin{subfigure}{0.31\textwidth}
        \includegraphics[width=\textwidth]{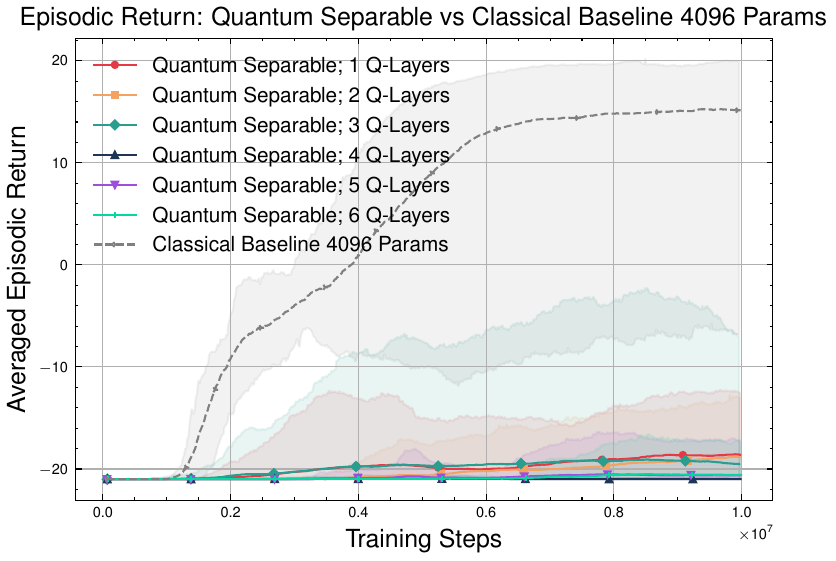}
        \caption{ }
    \end{subfigure}
    \hfill
    \begin{subfigure}{0.32\textwidth}
        \includegraphics[width=\textwidth]{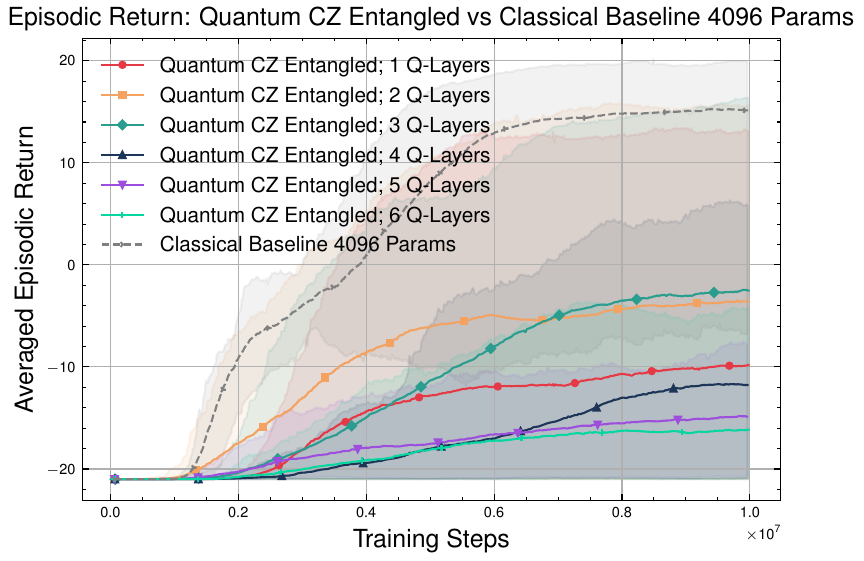}
        \caption{ }
    \end{subfigure}
    \hfill
    \begin{subfigure}{0.33\textwidth}
        \includegraphics[width=\textwidth]{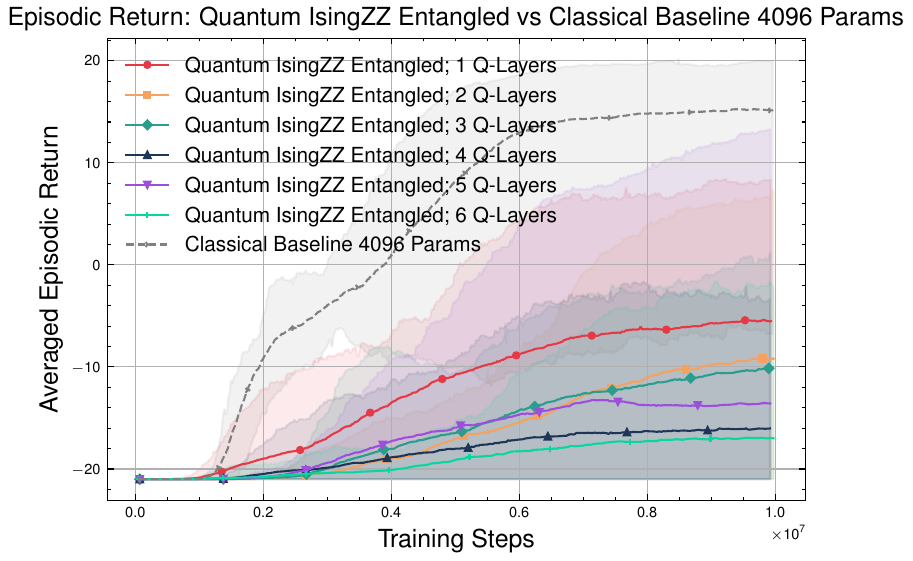}
        \caption{ }
    \end{subfigure}
    \vspace{0.2em}

    \caption{
        Comparing the (averaged) episodic performance of the three different quantum backbones (separable, CZ-entangled and IsingZZ-entangled) w.r.t. classical MLP with $128$, $256$, $336$ and $4096$ parameters. 
    }
    \label{fig:quant-vs-mlp-supp}
\end{figure*}

\clearpage
\bibliography{apssamp}
\end{document}